\documentclass[aps,prc,twocolumn,superscriptaddress,longbibliography]{revtex4-1}
\usepackage{epsfig,dsfont,amssymb,amsmath,amsthm,amsfonts,amsbsy,mathrsfs}
\usepackage{graphicx}
\usepackage{amsmath}
\usepackage{amssymb}%
\usepackage{dcolumn}
\usepackage{hyperref}
\usepackage[disable]{todonotes}

\newcommand{\be}{\begin{equation}}
\newcommand{\ee}{\end{equation}}
\newcommand{\ba}{\begin{eqnarray}}
\newcommand{\ea}{\end{eqnarray}}

\DeclareUnicodeCharacter{202F}{\,}
\begin{document}

\title{Accurate bulk properties of nuclei from $A = 2$ to $\infty$
  from potentials with $\Delta$ isobars}

\author{W.G. Jiang}
\affiliation{Department of Physics and Astronomy, University of Tennessee, Knoxville, TN 37996, USA}
\affiliation{Physics Division, Oak Ridge National Laboratory, Oak Ridge, TN 37831, USA}
\affiliation{Department of Physics, Chalmers University of Technology, SE-412 96 G\"oteborg, Sweden}

\author{A.~Ekstr\"om}
\affiliation{Department of Physics, Chalmers University of Technology, SE-412 96 G\"oteborg, Sweden}

\author{C.~Forss\'en}
\affiliation{Department of Physics, Chalmers University of Technology, SE-412 96 G\"oteborg, Sweden}

\author{G.~Hagen}
\affiliation{Physics Division, Oak Ridge National Laboratory, Oak Ridge, TN 37831, USA}
\affiliation{Department of Physics and Astronomy, University of Tennessee, Knoxville, TN 37996, USA}

\author{G.R. Jansen}
\affiliation{National Center for Computational Sciences, Oak Ridge National Laboratory, Oak Ridge, Tennessee 37831, USA}   
\affiliation{Physics Division, Oak Ridge National Laboratory, Oak Ridge, TN 37831, USA}
   
\author{T.~Papenbrock}
\affiliation{Department of Physics and Astronomy, University of Tennessee, Knoxville, TN 37996, USA}
\affiliation{Physics Division, Oak Ridge National Laboratory, Oak Ridge, TN 37831, USA}

\begin{abstract}
  We optimize $\Delta$-full nuclear interactions from chiral effective
  field theory. The low-energy constants of the contact potentials
  are constrained by two-body scattering phase shifts, and by
  properties of bound-state of $A=2$ to $4$ nucleon systems and
  nuclear matter. The pion-nucleon couplings are taken from a
  Roy-Steiner analysis. The resulting interactions yield accurate
  binding energies and radii for a range of nuclei from $A=16$ to
  $A=132$, and provide accurate equations of state for nuclear matter
  and realistic symmetry energies. Selected excited states are also
  in agreement with data.
\end{abstract}

\maketitle

\section{Introduction}
Ideas from chiral effective field theory (EFT) and the renormalization
group~\cite{vankolck1999,bogner2003,epelbaum2009,bogner2010,machleidt2011},
advances in computing power, and developments of many-body methods
that scale polynomially with mass number, have propelled {\it ab
  initio} calculations of atomic nuclei from light
\cite{barrett2013,carlson2015} to medium-mass isotopes
\cite{dickhoff2004,hergert2013b,soma2014,lahde2014,hagen2014,hergert2016,stroberg2017,morris2018,lonardoni2018}.
Such approaches are now starting to explore ever-increasing fractions
of the nuclear chart~\cite{holt2019}---a task once thought to be
reserved for computationally less expensive mean-field
methods~\cite{erler2012,agbemava2014,shen2019}. {\it Ab initio}
computations make controlled approximations that give a quantifiable
precision in the solution of the quantum many-body problem. All these
calculations are only as accurate as the employed nuclear
Hamiltonians.

In the past two decades, our understanding of the nuclear interactions
has evolved from phenomenological
models~\cite{wiringa1995,machleidt2001} to potentials whose
improvement is guided by ideas from
EFT~\cite{vankolck1994,epelbaum2009,machleidt2011}. The quest to link
such potentials to quantum chromodynamics, the microscopic theory of
the strong nuclear interaction, is
ongoing~\cite{barnea2015,contessi2017a,mcilroy2018,bansal2018}. Recent
developments in nuclear potentials from chiral EFT include (i) the
identification~\cite{reinert2018,wesolowski2019} of a redundant term
at next-to-next-to-next-to leading order, (ii) the systematic and
simultaneous optimization of nucleon-nucleon and three-nucleon
potentials~\cite{ekstrom2015,carlsson2016}, (iii) the construction of
local potentials for use with quantum Monte Carlo
methods~\cite{gezerlis2013,piarulli2015}, and (iv) the development of
high-order
interactions~\cite{entem2015,epelbaum2015,hebeler2015b,hebeler2020}.

In spite of these advances, nuclear potentials from chiral EFT have
long struggled to accurately reproduce bulk nuclear properties such as
charge radii and binding energies of finite nuclei, and the saturation
point and the symmetry energy of infinite nuclear matter. Notable
exceptions are the 1.8/2.0(EM) \cite{hebeler2011} and the
$NN$+$3N$(lnl) interactions \cite{soma2020} (which accurately
reproduce binding energies and spectra of selected nuclei up to tin
isotopes~\cite{morris2018} but yield too small radii), and the
NNLO$_{\text{sat}}$ potential~\cite{ekstrom2015} (which accurately
describes binding energies and radii up to calcium
isotopes~\cite{hagen2015} but is less accurate for spectra). The novel
family of interactions~\cite{huether2019} also seems promising but is
not much explored yet. While it is not clear what distinguishes these
particular interactions from their many peers that become inaccurate
beyond oxygen isotopes~\cite{binder2013b}, some key ingredients to
nuclear binding have been uncovered: Nuclear lattice EFT computations
revealed that non-locality is essential for low-momentum interactions
(with momentum cutoffs below about 600~MeV)~\cite{lu2019}, and that
elastic alpha-alpha scattering, for example, is very sensitive to the
degree of non-locality~\cite{elhatisari2016}. This casts some doubts
on the viability of soft local interactions. In our interpretation, these
results indicate that the finite size of nucleons plays an important
role in nuclear binding~\footnote{The hard core in local interactions
  also accounts for the finite nucleon size when large momentum
  cutoffs can be tolerated~\cite{wiringa1995}.}. The importance of
this length scale has been highlighted very recently by
\textcite{miller2020}. Consistent with this view is the finding that
the inclusion of $\Delta$ isobar degrees of freedom, i.e. excited
states of the nucleon that reflect its finite size, considerably
improve the saturation properties of chiral
potentials~\cite{ekstrom2017}. In addition, a recent statistical
analysis~\cite{ekstrom2019} reveals that the nuclear radius,
i.e. implicitly the nuclear saturation density, depends very
sensitively on the details of the chiral interaction. This suggests
that it might be profitable to include $\Delta$-isobar degrees of
freedom and nuclear matter properties into the construction and
optimization, respectively, of nuclear
potentials~\cite{simonis2017,drischler2019}.

In this paper we report on chiral potentials that accurately describe
bulk properties of finite nuclei and nuclear matter. We optimized
$\Delta$-full interactions and included nuclear matter properties as
calibration data.  We used the coupled-cluster (CC) method to compute
ground-state energies, charge radii, and spectra of nuclei up to tin,
plus the equation of state for nuclear matter.

\section{Optimization}
The $\Delta$-full interactions are based on
Refs.~\cite{vankolck1994,hemmert1998,kaiser1998,krebs2007,epelbaum2008},
and its specific form was used in Ref.~\cite{ekstrom2017}.  We
employed standard nonlocal regulator functions
$f(p)=\exp{(p/\Lambda)^{2n}}$ that act on relative momenta $p$, see,
e.g., Refs.~\cite{entem2003,epelbaum2002}. We constructed three
potentials: two of them with $\Lambda=450$~MeV (and power $n =3$ in
the regulator) at next-to-leading order (NLO) and
next-to-next-to-leading (NNLO), the other one at NNLO with a cutoff
$\Lambda=394$~MeV (and power $n=4$).  The softer interaction has a
momentum cutoff and regulator power exactly as the 1.8/2.0(EM)
potential. At NNLO there are 17 low-energy coefficients (LECs) that
parametrize the interaction. The pion-nucleon LECs $c_{1,2,3,4}$ were
held fixed during the optimization and taken as the central values
from the recent Roy-Steiner analysis~\cite{siemens2017}---see
Table~\ref{tab:LECs}. The LECs of the nucleon-nucleon and
three-nucleon potentials were simultaneously constrained by the
following data: low-energy nucleon-nucleon scattering data from the
Granada phase shift analysis~\cite{navarro2013} up to 200~MeV
scattering energy in the laboratory system; observables of few-nucleon
systems (with mass numbers $A\leq 4$) as listed in
Table~\ref{tab:light_nuclei}; the saturation energy and density, and
constraints on the symmetry energy and its slope of nuclear matter
from a lower bound on the neutron-matter energy~\cite{tews2017}. The
inclusion of the latter is in contrast to the construction of the
potential NNLO$_{\rm sat}$~\cite{ekstrom2015} which exhibits
deficiencies for neutron-rich nuclei and neutron matter.  The
minimization of the objective function was performed with the
algorithm POUNDerS~\cite{kortelainen2010}.  During this process we
periodically calculated selected medium-mass nuclei to further guide
the optimization. This allowed us to properly adjust the weights of
the nuclear matter properties in the objective function
\begin{equation}
f(\vec x)=w_1 \sum\limits_{p=1}^{N_p}r^2_p(\vec x) + w_2 \sum\limits_{q=1}^{N_q}r^2_q(\vec x)+w_3 \sum\limits_{s=1}^{N_s}r^2_s(\vec x) ,
\end{equation}
where $\vec x$ denotes the parameters of the interaction, $r_i(\vec
x)=( \mathcal{O}_{i}^\text{theo}(\vec x)-
\mathcal{O}_{i}^\text{exp})/\delta_{i}$ is the residual of observable
$\mathcal{O}_{i}$ with uncertainty $\delta_{i}$ determining its
weight, $r_p$, $r_q$ and $r_s$ are the residuals for phase shifts,
few-nucleon systems and nuclear matter respectively and $w_i$ are
their corresponding weights. Note that the weight of the $^2$H
quadrupole moment is increased to improve its description.

The inclusion of nuclear matter properties into the optimization
procedure is not without challenges. Here, we used coupled-cluster
calculations~\cite{hagen2013b}, which are based on a discrete lattice
in momentum space. Nondegenerate reference states are ``closed shell''
configurations, and we used periodic boundary conditions in position
space. Systems of 132 nucleons and 66 neutrons exhibit one of the
smallest finite-size corrections for symmetric nuclear matter and
neutron matter,
respectively~\cite{gandolfi2009,hagen2013b}. Unfortunately, such large
particle numbers are numerically too expensive to be used in the
optimization. However, systems of 28 nucleons and 14 neutrons exhibit
predictable differences (about 10\%) from systems consisting of 132
and 66 particles, respectively~\cite{hagen2013b}. This allowed us to
use the lower-precision computations with smaller system sizes in the
optimization. We checked periodically that our estimates for the
finite-size corrections were accurate.

To explore the resolution-scale dependence, we optimized the
interaction using two cutoffs, namely $\Lambda=450$ and $\Lambda=394$
~MeV at the NNLO level.  For $\Lambda=450$~MeV we also optimized an
NLO interaction. Table~\ref{tab:LECs} shows the optimized LECs values
for the three interactions of this work. In what follows, we label
these ``Gothenburg--Oak Ridge'' potentials as $\Delta
\rm{NLO}_{\rm{GO}}(450)$, $\Delta \rm{NNLO}_{\rm{GO}}(450)$ and
$\Delta \rm{NNLO}_{\rm{GO}}(394)$.
Most of the LECs of the newly constructed potentials are close to the
starting point of Ref.~\cite{ekstrom2017}, with a few exceptions. In
particular, the $c_D$ and $c_E$ for the short-ranged three-nucleon
forces have different signs for {$\Delta$}NNLO$_{\rm{GO}}$(450).

\begin{table}
\caption{
\label{tab:LECs}
Parameters for the new $\Delta$-full potentials with momentum cutoffs
$\Lambda=450$ and $394$~MeV. The constants $c_i$, $\tilde{C_i}$
and $C_i$ are in units of $\text{GeV}^{-1}$, $10^4\text{GeV}^{-2}$ and
$10^4\text{GeV}^{-4}$, respectively.  }
\begin{ruledtabular}
\begin{tabular}{l lll p{cm}}
LEC &{$\Delta$}NLO$_{\rm{GO}}$(450)&{$\Delta$}NNLO$_{\rm{GO}}$(450) &{$\Delta$}NNLO$_{\rm{GO}}$(394)\\
\colrule
$c_{1}$                        &  $-$&  $-$0.74     &  $-$0.74        \\
$c_{2}$                        &  $-$&  $-$0.49     &  $-$0.49        \\
$c_{3}$                        &  $-$&  $-$0.65     &  $-$0.65        \\
$c_{4}$                       &  $-$ &  $+$0.96     &  $+$0.96       \\
$\tilde{C}^{(nn)}_{^1S_{0}}$  &$-$0.314882 & $-$0.339887  &$-$0.338746 \\
$\tilde{C}^{(np)}_{^1S_{0}}$  &$-$0.315639 & $-$0.340114  &$-$0.339250 \\
$\tilde{C}^{(pp)}_{^1S_{0}}$  &$-$0.314300 & $-$0.339111  &$-$0.338142 \\
$\tilde{C}_{^3S_{1}}$           &$-$0.234132  & $-$0.253950  &$-$0.259839 \\
$C_{^1S_{0}}$                     &$+$2.521650     & $+$2.526636  &$+$2.505389 \\
$C_{^3S_{1}}$                     &$+$1.025459     & $+$0.964990  &$+$1.002189 \\
$C_{^1P_{1}}$                      &$+$0.152206     & $-$0.219498  &$-$0.387960 \\
$C_{^3P_{0}}$                     &$+$0.671880     & $+$0.671908  &$+$0.700499 \\
$C_{^3P_{1}}$                      &$-$0.955644     & $-$0.915398  &$-$0.964856 \\
$C_{^3P_{2}}$                      &$-$0.824639     & $-$0.895405  &$-$0.883122 \\
$C_{^3S_{1}-^3D_1}$           &$+$0.451306  & $+$0.445743  &$+$0.452523 \\
$c_{D}$                       &$-$ & $-$0.454     &$+$0.081     \\
$c_{E}$                       &$-$ & $-$0.186     &$-$0.002      \\
\end{tabular}
\end{ruledtabular}
\end{table}

\begin{figure}
  \includegraphics[width=0.45\textwidth]{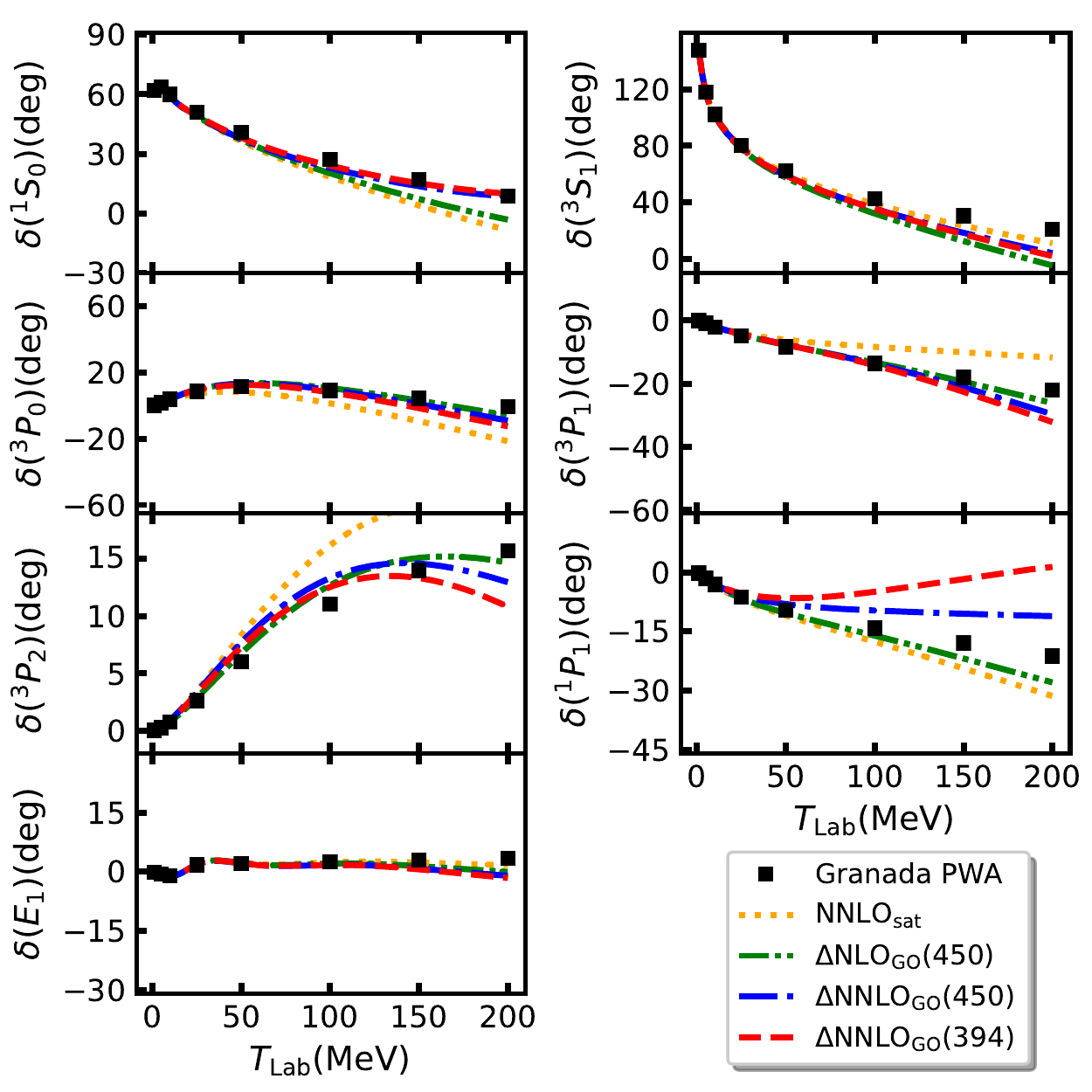}
  \caption{(Color online) Computed neutron-proton phase shifts for the
    contact partial waves with the {$\Delta$}NLO$_{\rm{GO}}$ and
    {$\Delta$}NNLO$_{\rm{GO}}$ potentials (dashed),
    $\text{NNLO}_{\text{sat}}$~\cite{ekstrom2015} (red, dotted) and
    compared with the Granada phase shift analysis \cite{navarro2013}
    (black squares)
  \label{fig:phase_shift1}}
\end{figure}

\begin{figure}
  \includegraphics[width=0.45\textwidth]{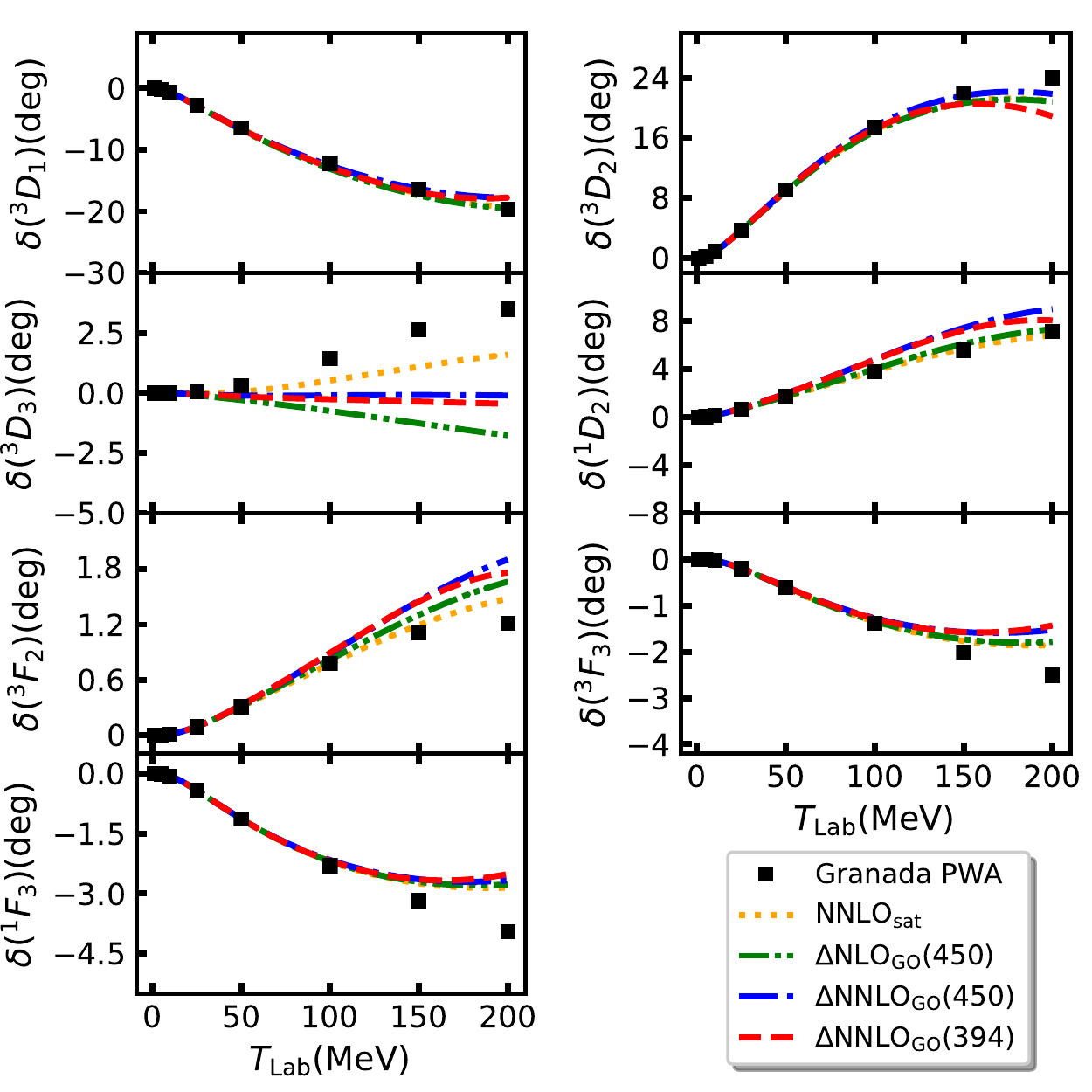}
  \caption{(Color online) Computed neutron-proton phase shifts for the
    selected peripheral partial waves with the
    {$\Delta$}NLO$_{\rm{GO}}$ and {$\Delta$}NNLO$_{\rm{GO}}$
    potentials (dashed), $\text{NNLO}_{\text{sat}}$~\cite{ekstrom2015}
    (red, dotted) and compared with the Granada phase shift analysis
    \cite{navarro2013} (black squares)
  \label{fig:phase_shift2}}
\end{figure}

\begin{figure}
  \includegraphics[width=0.45\textwidth]{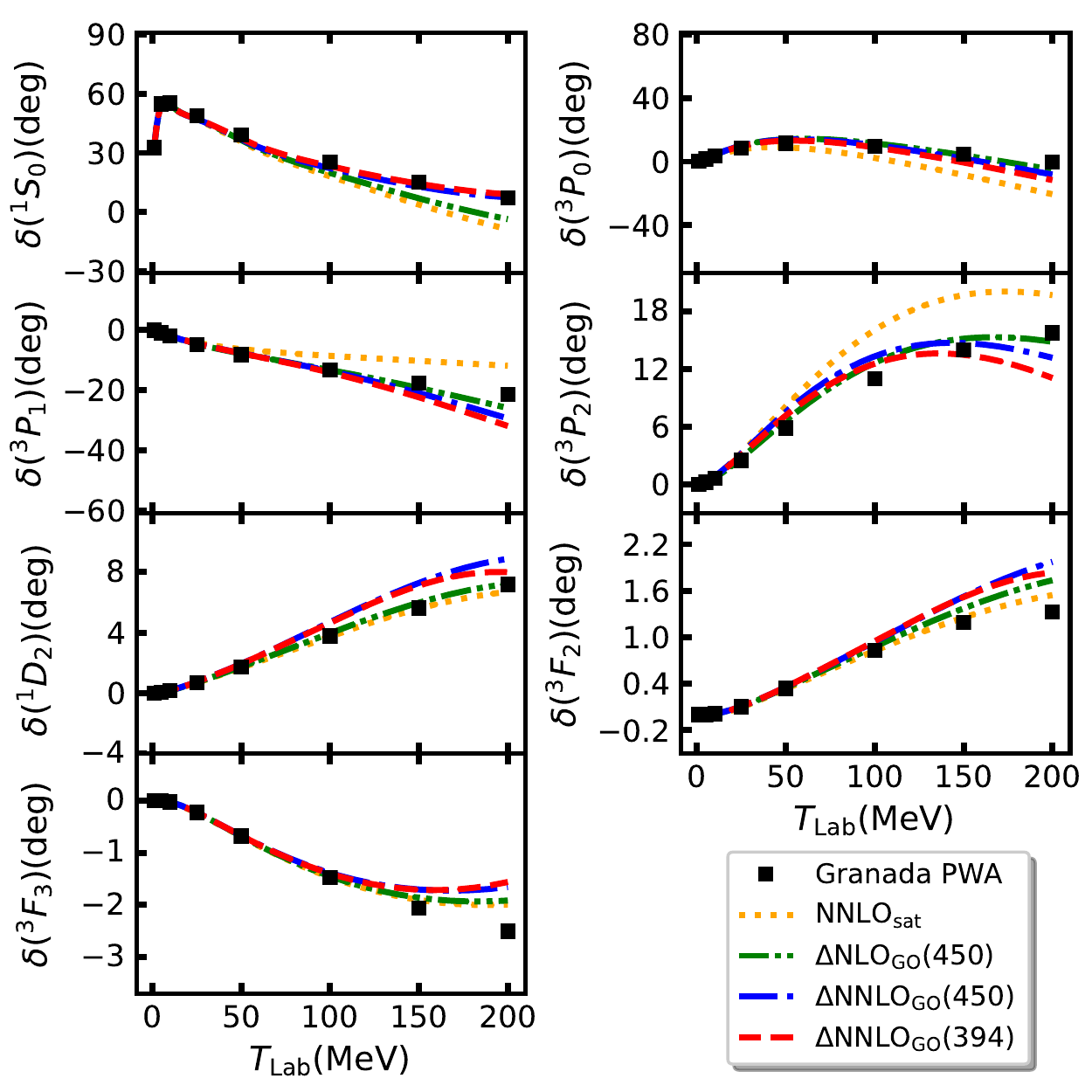}
  \caption{(Color online) Computed proton-proton phase shifts for the
    contact and selected peripheral partial waves with new
    {$\Delta$}NLO$_{\rm{GO}}$ and {$\Delta$}NNLO$_{\rm{GO}}$
    potentials (dashed), $\text{NNLO}_{\text{sat}}$~\cite{ekstrom2015}
    (red, dotted) and compared with the Granada phase shift analysis
    \cite{navarro2013} (black squares)
  \label{fig:phase_shift3}}
\end{figure}

Figures \ref{fig:phase_shift1}, \ref{fig:phase_shift2} and
\ref{fig:phase_shift3} show the phase shifts of the new potentials
for a few representative neutron-proton channels and compares them to
the Granada partial wave analysis~\cite{navarro2013}. Overall, the
phase shifts of {$\Delta$}NNLO$_{\rm{GO}}$ are improved compared to
$\text{NNLO}_{\text{sat}}$, and they are close to the data for
laboratory energies below about 125~MeV. We note that the phase shifts
of $\Delta$NLO$_{\rm{GO}}$(450) are within the uncertainty estimates
expected at this order~\cite{epelbaum2015,ekstrom2017,furnstahl2015a}.
One should see that some partial waves such as $^1\text{P}_1$ and
$^3\text{D}_3$ are less accurate than others and might need higher
orders of the potentials to give a better description.

Table~\ref{tab:light_nuclei} summarizes the results of bound-state
observables for light nuclei with $A \leq 4$. The theoretical results
were obtained with the no-core shell model (NCSM)~\cite{barrett2013}
in translationally invariant Jacobi
coordinates~\cite{navratil2000b}. These calculations used an
oscillator frequency of $\hbar\omega=36 (28)$~MeV for $\Lambda = 450
(394)$~MeV in model spaces consisting of $N_{\rm{max}}=40$ and
$N_{\rm{max}}=20$ oscillator shells for $A=3$ and $A=4$ nuclei,
respectively. They are converged within these model spaces. The charge
radii shown in Table~\ref{tab:light_nuclei} are obtained from the
computed point-proton radii with standard nucleon-size and
relativistic corrections, see, e.g., Ref~\cite{ekstrom2015}. The two
$\Delta$NNLO$_{\rm{GO}}$ potentials reproduce the experimental
energies within less than 0.5\% and other observables within 2\%.  We
note that it was important to include the deuteron quadrupole moment
as a calibration datum in the optimization. The value for the
quadrupole moment that we targeted, i.e. $Q=0.27 ~\mathrm{efm}^2$, was
obtained from a theoretical calculation based on the high-precision
meson-exchange $NN$ model CD-Bonn.

\begin{table}
\caption{
\label{tab:light_nuclei}
Binding energies ($E$) in MeV, charge radii ($R_{ch}$) in fm, for
$^{2,3}$H and $^{3,4}$He computed with the interactions developed in
this work and compared to data~\cite{angeli2013,wang2017}. The
quadrupole moment ($Q$) in efm$^2$ for the ground state of
$^{2}\rm{H}$ is also shown. The $D$-state probability is 3.06\%,
3.12\%, and 2.97 \% for the three interactions in column order.}
\begin{ruledtabular}
\begin{tabular}{c D{.}{.}{2.4}  D{.}{.}{2.4} D{.}{.}{2.4} D{.}{.}{2.4} }
& \multicolumn{1}{c}{$\Delta{\rm{NLO}}_{\rm{GO}}$} & \multicolumn{1}{c}{$\Delta{\rm{NNLO}}_{\rm{GO}}$} & \multicolumn{1}{c}{$\Delta {\rm{NNLO}}_{\rm{GO}}$} & \multicolumn{1}{c}{Expt.} \\
& \multicolumn{1}{c}{(450)} & \multicolumn{1}{c}{(450)} & \multicolumn{1}{c}{(394)} & \\
\colrule
$E$($^2$H)            & 2.2586  & 2.2358   & 2.2298 & 2.2245  \\
$R_\mathrm{ch}$($^2$H)       & 2.1511 & 2.1509   & 2.1531 & 2.1421  \\
$Q$($^2$H)            & 0.2680 & 0.2675   & 0.2674 & 0.27 \footnote{CD-Bonn value according to Ref.~\cite{machleidt2011}. See the text for details.}    \\
$E$($^3$H)            & 8.4803 & 8.4809   & 8.4812 & 8.4818  \\
$R_\mathrm{ch}$($^3$H)       & 1.7928 & 1.7801   & 1.7833 & 1.7591  \\
$E$($^3$He)           & 7.7495 & 7.7162   & 7.7245 & 7.7180  \\
$R_\mathrm{ch}$($^3$He)      & 1.9954 & 2.0036   & 1.9946 & 1.9661  \\
$E$($^4$He)           & 28.3945 & 28.2975  & 28.3028 & 28.2957 \\
$R_\mathrm{ch}$($^4$He)      & 1.7099 & 1.6960   & 1.6919 & 1.6775  \\
\end{tabular}
\end{ruledtabular}
\end{table}

\begin{figure}
  \includegraphics[width=0.45\textwidth]{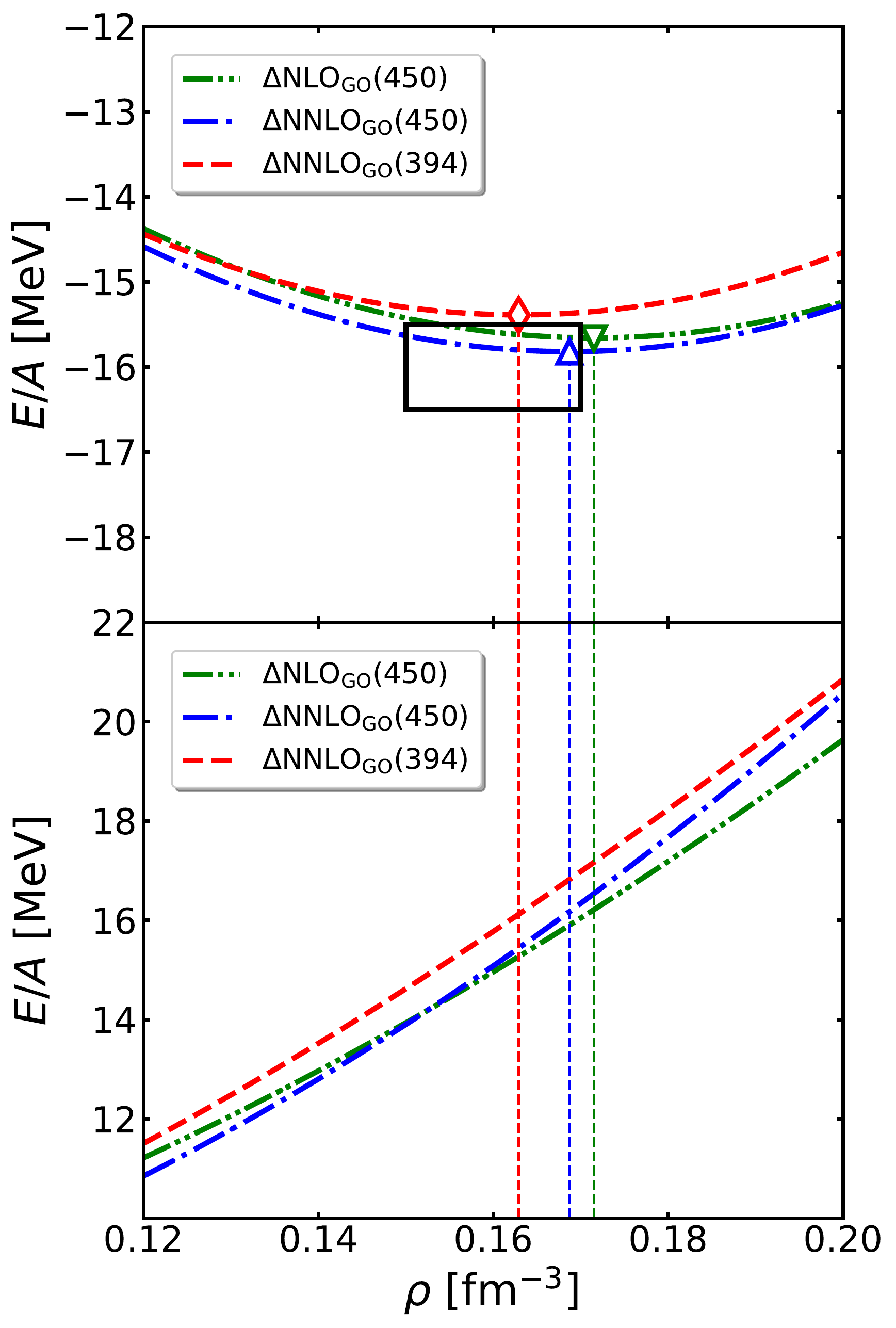}
  \caption{(Color online) Energy per nucleon (in MeV) for symmetric
    nuclear matter (top) and pure neutron matter (bottom) with $\Delta
    \rm{NLO}_{\rm{GO}}(450)$, $\Delta \rm{NNLO}_{\rm{GO}}(450)$ and
    $\Delta \rm{NNLO}_{\rm{GO}}(394)$. The black rectangle indicates
    the empirical saturation point. }
  \label{fig:nuclear_matter}
\end{figure}

Figure~\ref{fig:nuclear_matter} shows the energy per nucleon in
symmetric nuclear matter (top) and pure neutron matter (bottom) as a
function of density, using 132 nucleons and 66 neutrons, respectively.
The black rectangle indicates the empirical saturation region with
$E/A=-16 \pm 0.5$~MeV and
$\rho = 0.16\pm 0.01$~fm$^{-3}$~\cite{bender2003,hebeler2011}.  The CC
calculations were performed in the CCD(T) approximation, i.e. with
2$p$-2$h$ excitations and perturbative 3$p$-3$h$ corrections as done
in Refs.~\cite{ekstrom2017, hagen2013b}.  We find the saturation
density $\rho_{0}= 0.169 \ \rm{fm}^{-3}$, the symmetry energy
$S_{0}=32.0\ \rm{MeV}$ and its slope $L=65.2$ for the
$\Delta \rm{NNLO}_{\rm{GO}}(450)$ potential, and
$\rho_{0}= 0.163 \ \rm{fm}^{-3}$, $S_{0}=31.5\ \rm{MeV}$ and $L=58.4$
for $\Delta \rm{NNLO}_{\rm{GO}}(394)$. These nuclear-matter properties
are more accurate than those reported in Ref.~\cite{ekstrom2017}, and
in good agreement with the recent predictions of the symmetry energy
and its slope obtained using Bayesian machine learning techniques to
quantify EFT uncertainties of the nuclear matter equation of
state~\cite{drischler2020}.

\section{Results}
Our CC computations of heavier nuclei started from a spherical
Hartree-Fock state built from a model-space consisting of 15 major
oscillator shells with frequency $\hbar \omega=16$~MeV. The
three-nucleon force had the additional energy cut of
$E_\mathrm{3max}=16 \hbar\omega$.  Our calculations employed the
coupled-cluster singles, doubles, and leading triples approximation
CCSDT-1~\cite{lee1984}.  No further truncations were imposed on the
three-particle-three-hole amplitudes.  This computational achievement
was enabled by the Nuclear Tensor Contraction
Library~\cite{ntclwebsite} ---a domain-specific library, dedicated to
the sparse tensor contractions that dominate in coupled-cluster
method---that is developed to run at scale on Summit, the
U.S. Department of Energy's 200 petaflop supercomputer operated by the
Oak Ridge Leadership Computing Facility (OLCF) at Oak Ridge National
Laboratory.

Table~\ref{tab:heavier_nuclei} shows the binding energies of selected
closed-shell nuclei up to $^{132}$Sn.  We found that $\Delta
\rm{NNLO}_{\rm{GO}}(394)$ converges faster, especially for heavier
nuclei, than $\Delta \rm{NNLO}_{\rm{GO}}(450)$.  This is expected due
to its lower momentum cutoff. A dash indicates that the employed model
space was too small to achieve reasonably converged energies.
Uncertainties from the coupled-cluster method (about 30\% of the
difference between doubles and triples energies) and from the model
space are given in subsequent parenthesis, respectively. For lighter
nuclei, the uncertainties from the model-space are omitted because
they are much smaller than those of the method. The model-space
uncertainties combine the truncation of single-particle model space
and the employed $E_\mathrm{3max}=16$ cut of the three-nucleon
interaction. Reference~\cite{morris2018} found that for the
1.8/2.0(EM) interaction the binding energy of $^{100}$Sn changes by
less than 1\% by increasing $E_\mathrm{3 max}$ from $E_\mathrm{3
  max}=16$ to $E_\mathrm{3max}=18$. This finding guided our estimated
model-space uncertainty as the 1.8/2.0(EM) has identical three-nucleon
regulator and momentum cutoff as the $\Delta \rm{NNLO}_{\rm{GO}}(394)$
potential. It is non-trivial to estimate the EFT truncation errors for
bound nuclear states since the relevant momentum scale is unknown and
the lack of a spin-orbit (LS) force at leading order (LO) give energy
degeneracies that hamper CC calculations of non-LS-closed
nuclei. Nevertheless, based on the observed order-by-order convergence
in Ref.~\cite{ekstrom2017}, we estimate the EFT truncation errors for
the $\Delta \rm{NNLO}_{\rm{GO}}$ interactions to 1 MeV and 7 MeV in
the ground state energies of oxygen and calcium isotopes
respectively. We also expect the truncation error to be the dominating
source of uncertainty in heavier nuclei. For symmetric nuclear matter
and pure neutron matter we expect a truncation error of $\pm 1$ MeV
and $\pm 2$ MeV per nucleon, respectively, with $\Delta
\rm{NNLO}_{\rm{GO}}$, see Ref.~\cite{ekstrom2017} for details.

\begin{table}
\caption{
\label{tab:heavier_nuclei}
Binding energies (in MeV) for selected nuclei with the new interaction
using CCSDT-1 and compared to data. }
\begin{ruledtabular}
\begin{tabular}{c lllr }
& \multicolumn{1}{c}{$\Delta{\rm{NLO}}_{\rm{GO}}$} & \multicolumn{1}{c}{$\Delta{\rm{NNLO}}_{\rm{GO}}$} & \multicolumn{1}{c}{$\Delta {\rm{NNLO}}_{\rm{GO}}$} & \multicolumn{1}{c}{Expt.} \\
& \multicolumn{1}{c}{(450)} & \multicolumn{1}{c}{(450)} & \multicolumn{1}{c}{(394)} & \\
\colrule
$^{16}$O   & 128.2  & 128.1(23)          & \;\;127.5(19)       & 127.62  \\
$^{24}$O   & 165    & 170\,\;\;(3)      & \;\;169\;\;\,(3)         & 168.96  \\
$^{40}$Ca  & 341    & 348\;\;\,(7)(1)   & \;\;346\;\;\,(6)         & 342.05  \\
$^{48}$Ca  & 410    & 422\;\;\,(9)(4)   & \;\;420\;\;\,(7)         & 416.00     \\
$^{78}$Ni  & -      & 631\;\;\,(14)(20)  & \;\;639\;\;\,(11)(4)     & 641.55  \\
$^{90}$Zr  & -      & \;\;\,\;\;\,-     & \;\;782\;\;\,(14)(6)     & 783.90   \\  
$^{100}$Sn & -      & \;\;\,\;\;\,-     & \;\;818\;\;\,(16)(7)     & 825.30 \\
$^{132}$Sn & -      & \;\;\,\;\;\,-     & 1043\,\;\;(20)(30)    & 1102.84   \\
\end{tabular}
\end{ruledtabular}
\end{table}

\begin{figure}
  \includegraphics[width=0.45\textwidth]{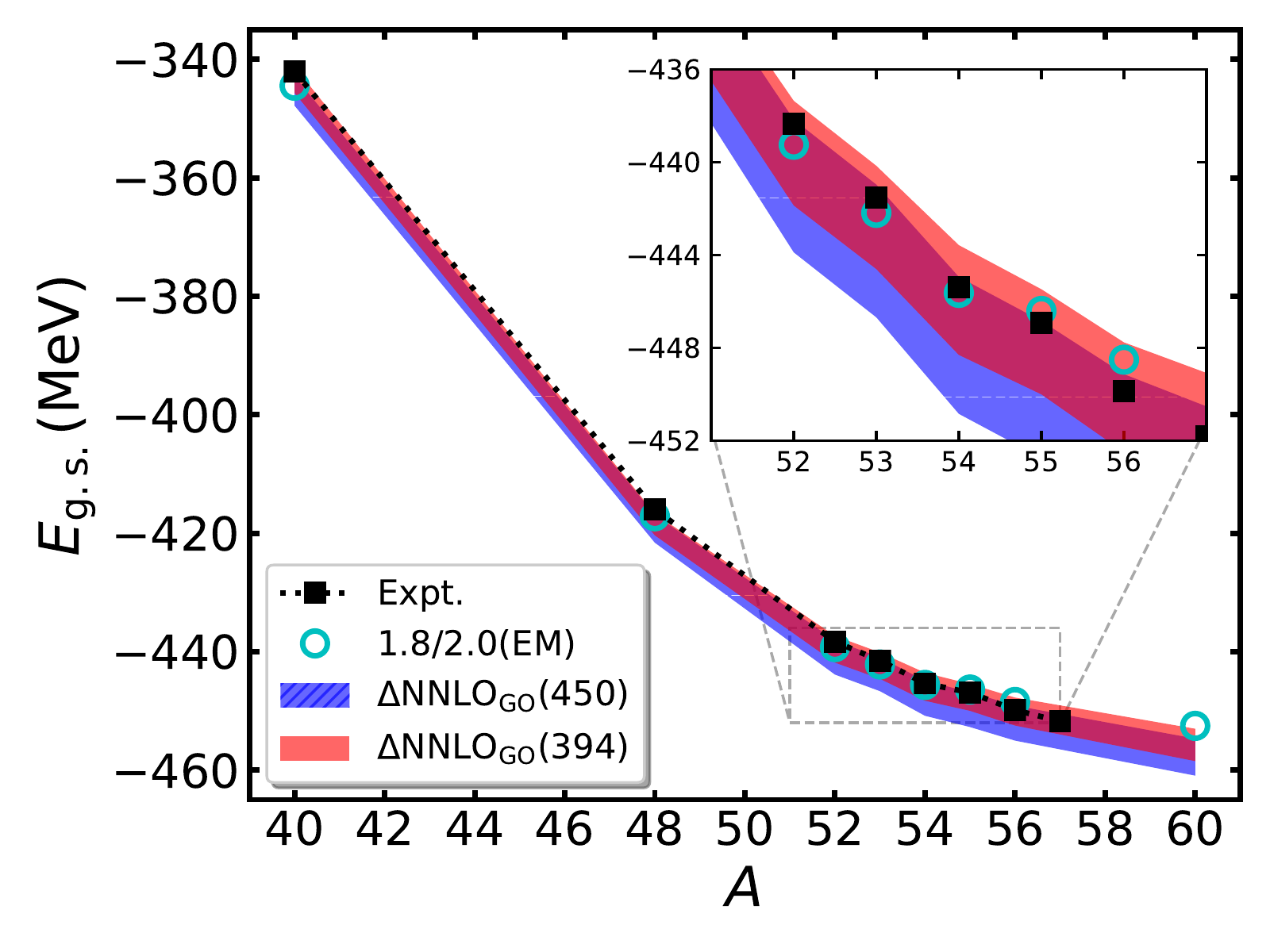}
  \caption{(Color online) The ground-state energies of calcium
    isotopes obtained with $\Delta \rm{NNLO}_{\rm{GO}}$ and
    1.8/2.0(EM) interaction compared with experiment (data of
    $^{55-57}$Ca are taken from Ref.~\cite{michimasa2018}).  }
  \label{fig:ca_chain_gs}
\end{figure}

Figure \ref{fig:ca_chain_gs} shows the ground-state energies of
selected calcium isotopes and compares them to the 1.8/2.0(EM)
interaction~\cite{strobergpc} and to data. The lower borders of the
bands are results from CCSDT-1 while the upper borders are from
$\Lambda$-CCSD(T)~\cite{taube2008} which treats triples excitations
perturbatively. For $^{53,55}$Ca we employed the particle-attached
equation-of-motion coupled-cluster (EOM-CC) method with perturbative
three-particle-two-hole excitations from Ref.~\cite{morris2018}, while
for $^{56}$Ca we employed the two-particle attached EOM-CC method from
Refs.~\cite{jansen2011,jansen2012}. All interactions accurately
describe isotopes from $^{40}$Ca to $^{56}$Ca.  We note that the
ground-state of $^{60}$Ca is bound by about 10~MeV with respect to
$^{54}$Ca, consistent with recent data
~\cite{michimasa2018,tarasov2018,cortes2020}.

\begin{figure}
  \includegraphics[width=0.45\textwidth]{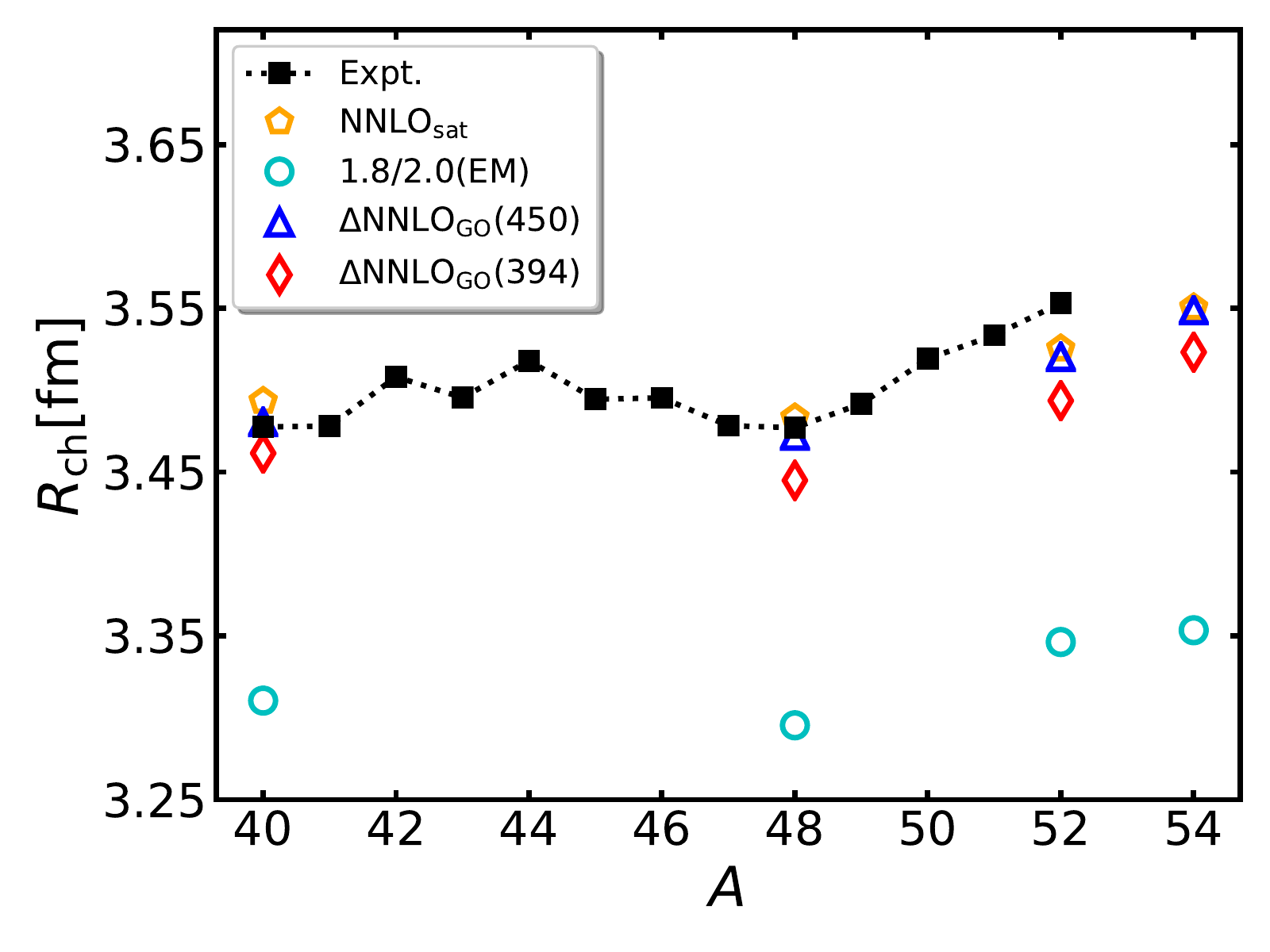}
  \caption{(Color online) Charge radii of calcium isotopes obtained
    with with $\Delta \rm{NNLO}_{\rm{GO}}$ and 1.8/2.0(EM) interaction
    compared with experiment. }
  \label{fig:ca_chain_rch}
\end{figure}

Figure~\ref{fig:ca_chain_rch} shows that the interactions of this work
yield significantly larger charge radii than the 1.8/2.0(EM)
potential. Nevertheless, the $\Delta$NNLO$_{\rm GO}$ potentials still
fail to explain the unusually large charge radii of neutron-rich
calcium isotopes. For speculations about the origin of these large
charge radii in calcium isotopes we refer the reader to
Ref.~\cite{garciaruiz2016}.

We employed EOM-CC methods with singles, doubles, and leading-order
triples excitations, EOM-CCSDT-1~\cite{watts1995}, to compute excited
states of $^{16,22,24}$O, and $^{48}$Ca. The EOM-CCSDT-1 method is
computationally demanding, and we therefore limited the number of
three-particle-three-hole excitations by employing the energy cut
$\tilde{E}_{pqr}=\tilde{e}_p+\tilde{e}_q+\tilde{e}_r<\tilde{E}_{\rm
  3max}$, where $ \tilde{e}_p=|N_p-N_{F}|$ is the energy difference of
the single-particle energies with respect to the Fermi surface
$N_F$. This energy cut improved the convergence of the EOM-CCSDT-1
method with respect to the number of three-particle-three-hole
excitations~\cite{miorelli2018a,gysbers2019}. In this work we also
employed the method developed in Ref.~\cite{gysbers2019} to correct
perturbatively for three-particle-three-hole excitations above
$\tilde{E}_{\rm 3max}$. Here we employed the energy cut
$\tilde{E}_{\rm 3max} = 6$ which was sufficient to converge all
excited states to within approximately 100~keV. The results are summarized in
Table~\ref{tab:spectra} with estimated error bars. The
$\Delta$NNLO$_{\rm GO}$(450) potential only exhibits marginal
agreement with the data for $^{22}$O. In contrast, all potentials
accurately reproduce the first $3^-$ state of $^{16}$O which reflects
that the charge radius is well reproduced in this nucleus (see
Ref.~\cite{ekstrom2015} for a more detailed discussion on this
point). The uncertainties are estimated based on
Refs.~\cite{hagen2016a,gysbers2019}. Here excited states were computed
in different truncations within the EOM-CC approach, and it was found
that the triples correction to the excited states were about 20\% of
the EOM-CCSD correlation energy. Using this, we give a conservative
error estimate in the first parenthesis which amounts to 6\% and 15\%
of the total excitation energies for the interactions with cutoffs
394~MeV and 450~MeV, respectively. The uncertainties from the
truncated model space are given in the second parenthesis.

\begin{table}
\caption{
\label{tab:spectra}
Energies (in MeV) of selected excited states for different nuclei
using $\Delta \rm{NNLO}_{\rm{GO}}(450)$ and $\Delta
\rm{NNLO}_{\rm{GO}}(394)$ with EOM-CCSDT-1 and compared to
experiment. The uncertainties reflect estimated equation-of-motion
coupled-cluster and model-space truncation errors, respectively.}
\begin{ruledtabular}
\begin{tabular}{c ccc  p{cm}}
 &\text{$\Delta$}NNLO$_{\rm{GO}}$(450) & \text{$\Delta$}NNLO$_{\rm{GO}}$(394) & Expt. \\
\colrule
$^{16}$O $3^-_1$        & 6(1)       & 5.6(3)(1)  & 6.13  \\
$^{22}$O $2^+_1$        & 2.2(3)(1)  & 3.0(2)(1)  & 3.20  \\
$^{24}$O $2^+_1$        & 3.4(5)(2)  & 3.9(2)(1)  & 4.79  \\
$^{48}$Ca $2^+_1$       & 3.5(5)(2)  & 4.1(2)(1)  & 3.83  \\
\end{tabular}
\end{ruledtabular}
\end{table}
Finally we note that the potentials developed in this work have
recently been applied to several other open-shell and deformed nuclei
such as $^{29}$F~\cite{bagchi2020}, $^{40}$Ar~\cite{payne2019}, and
neon and magnesium isotopes~\cite{novario2020}.

\section{Summary}
We developed chiral interactions with $\Delta$ degrees of freedom by
calibrating LECs to reproduce nucleon-nucleon scattering phase shifts,
bound-state observables of few-nucleon systems, and properties of
infinite nuclear matter. The resulting $\Delta \rm{NNLO}_{\rm{GO}}$
potentials yield accurate (within about 2\%) binding energies of
nuclei up to mass numbers $A=132$, and improved radii for medium-mass
nuclei. The description of neutron-rich calcium isotopes is improved
by including the symmetry energy in the optimization. Selected excited
states are also accurately reproduced.  This shows that key nuclear
properties can be obtained by chiral interactions at
next-to-next-to-leading order.

\begin{acknowledgments}
  We thank Titus Morris and Ragnar Stroberg for fruitful discussions,
  in addition we also thank Ragnar Stroberg for providing us with the
  1.8/2.0(EM) results for calcium isotopes.  Weiguang Jiang
  acknowledges support as an FRIB-CSC Fellow. This work was supported
  by the Office of Nuclear Physics, U.S. Department of Energy,
  under grants de-sc0018223 (SciDAC-4 NUCLEI collaboration), DE-
  FG02-96ER40963, and by the Field Work Proposal ERKBP72 at Oak Ridge
  National Laboratory (ORNL), the European Research Council (ERC)
  under the European Unions Horizon 2020 research and innovation
  programme (Grant agreement No. 758027), the Swedish Research Council
  (Grant No. 2017-04234), and the Swedish Foundation for International
  Cooperation in Research and Higher Education (STINT, IG2012-5158).
  Computer time was provided by the Innovative and Novel Computational
  Impact onTheory and Experiment (INCITE) program. This research used
  resources of the Oak Ridge Leadership Computing Facility and of the
  Compute and Data Environment for Science (CADES) located at ORNL,
  which is supported by the Office of Science of the Department of
  Energy under Contract No. DE AC05-00OR22725.
\end{acknowledgments}

\end{document}